# Experimental Investigation of Variations in Polycrystalline $Hf_{0.5}Zr_{0.5}O_2$ (HZO)-based MFIM

Tae Ryong Kim, Revanth Koduru, Zehao Lin, Graduate Student Member, *IEEE*, Peide. D. Ye, Fellow, *IEEE*, and Sumeet Kumar Gupta, Senior *Member, IEEE*

*Abstract*—Device-to-device variations in ferroelectric (FE) hafnium oxide ($HfO_2$)-based devices pose a crucial challenge that limits the otherwise promising capabilities of this technology. Although previous simulation-based studies have identified polarization (*P*) domain nucleation and polycrystallinity as key contributors to variations in $HfO_2$, experimental validation remains limited. Here, we experimentally investigate variations in remanent polarization ($P_R$) of $Hf_{0.5}Zr_{0.5}O_2$ (HZO)-based metal-ferroelectric-insulator-metal (MFIM) capacitors across different set voltages ($V_{SET}$) and FE thicknesses ($T_{FE}$). Our measurements reveal a non-monotonic behavior of the standard deviation of $P_R$ with $V_{SET}$ peaking around coercive voltage ($V_C$), which is consistent with previous simulation-based predictions. In the low- and high-$V_{SET}$ regions, $P_R$ variations are primarily dictated by saturation polarization ($P_S$) variations, mainly originating from charge trap effects at the interface between the FE-dielectric (DE) layer and the polycrystallinity of FE. On the other hand, in the mid-$V_{SET}$ region peak, the $P_R$ variations are attributed to the $V_C$ variation, which comes from a combined effect of multi-domain (MD) *P* switching and polycrystallinity. Notably, sharp *P* switching associated with domain nucleation amplifies the variations, resulting in a peak of $P_R$ variations in this $V_{SET}$ range. Further, we observe that as HZO thickness ($T_{FE}$) is scaled, the non-monotonicity in variations with $V_{SET}$ is reduced, primarily due to reduced domain nucleation and smaller grain sizes. We experimentally establish a strong correlation of $P_R$ with $P_S$ in the low- and high-$V_{SET}$ regions and with $V_C$ in the mid-$V_{SET}$ region across various $T_{FE}$. Finally, our experimental findings are corroborated with simulations using a 3D phase-field model.

## I. INTRODUCTION

FERROELECTRIC (FE) materials are considered promising candidates for next-generation electronics due to their spontaneous non-zero polarization (*P*) and other unique characteristics [1], [2], [3], [4]. The discovery of ferroelectricity in doped hafnium oxide ($HfO_2$) has triggered immense attraction due to its CMOS compatibility [5], [6]. This has led to the integration of $HfO_2$ in various devices, such as ferroelectric RAM (FERAM) [7], [8], ferroelectric field effect transistors (FEFETs) [6], [9], and ferroelectric tunnel junctions (FTJs) [10], [11] and their use in a range of applications [1], [2], [3], [4]. Among various dopants for $HfO_2$ (Si [12], Al [13], Y [14], La [15], etc.) [16], Zirconium (Zr) stands out for its lower annealing temperature (400-500°C) to exhibit FE properties [17], [18]. A low thermal budget facilitates the integration of Zr-doped $HfO_2$ ($Hf_{1-x}Zr_xO_2$ or HZO) with front-end-of-line (FEOL) conventional CMOS devices without introducing additional stress [18].

Despite their potential, several challenges hinder the commercialization of FE HZO devices [19], [20], [21], [22], [23], [24], [25]. Among those, device-to-device variations significantly impact the performance of HZO-based systems [21], [22], [23], [24], [25], [26]. Specifically, in the context of multi-state storage (multiple intermediate polarizations) [27], [28], [29], where the state of FE-based devices is determined by applied $V_{SET}$ (or $V_{RESET}$), device-to-device variations are shown to be significantly dependent on $V_{SET}$ (or $V_{RESET}$) [24], [25]. For example, the read current [24] and threshold voltage [25] of FEFET, determined by the remanent polarization ($P_R$) of the FE layer in the gate stack, has been experimentally shown to display non-monotonic dependence on $V_{SET}$ ($V_{RESET}$).

Several factors have been suggested as causes of *P* switching variations in HZO-based metal-ferroelectric-insulator-metal (MFIM), including polycrystallinity [30], [31], [32] multi-domain (MD) *P* switching [33], [34], [35], and the charge trap effects at the FE-insulator or dielectric (DE) interface [36], [37], [38]. Multiple simulation-based studies have highlighted the role of polycrystallinity and MD *P* switching in governing the trends in the variations of coercive voltage ($V_C$) and remanent polarization ($P_R$) from the standpoint of HZO-based MFIM capacitors [34], [35]. The variations in *P* switching characteristics in polycrystalline HZO devices are attributed to the dependence of $P_R$ and $V_C$ on the crystal orientation angle ($\theta$) of the grains [34] as well as the phase distributions [32], [39], [40]. Multi-domain (MD) *P* switching, prevalent in HZO due to

This work was supported in part by NSF and by the Center for the Co-Design of Cognitive Systems (COCOSYS), one of the seven centers in JUMP 2.0, funded by SRC and DARPA.

Tae Ryong Kim, Revanth Koduru, and Sumeet Kumar Gupta are with the Elmore Family School of Electrical and Computer Engineering, Purdue University, West Lafayette, IN 47907 USA (e-mail: kim3804@purdue.edu; kodurur@purdue.edu; guptask@purdue.edu).

Zehao Lin and Peide. D. Ye are the Elmore Family School of Electrical and Computer Engineering and Birck Nanotechnology Center, Purdue University, West Lafayette, IN 47907 USA. (e-mail: lin1174@purdue.edu; yep@purdue.edu).

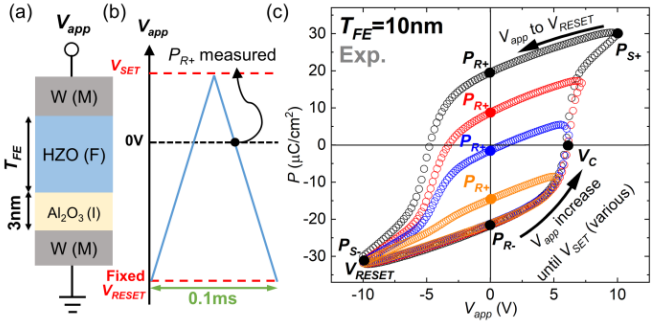
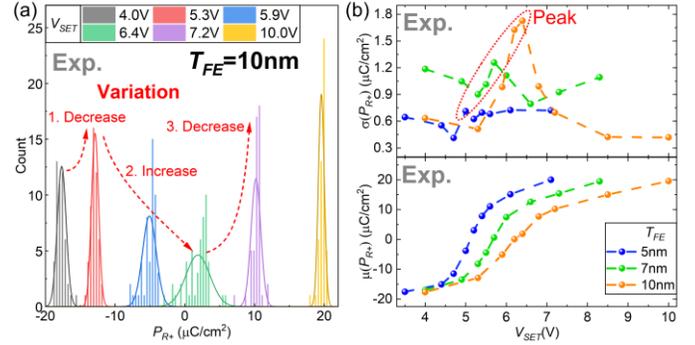

Fig. 1. (a) The schematic of the fabricated HZO-based metal-ferroelectric-insulator-metal (MFIM) capacitor (b) the waveform of 10kHz applied voltage ($V_{app}$) with positive remanent polarization ($P_{R+}$) measurement point and (c) measured polarization-voltage (*P-V*) curves averaged over 50 MFIM capacitors with 10nm of ferroelectric layer thickness ($T_{FE}$) for different set voltage ($V_{SET}$)

Fig. 2. (a) Measured $P_{R+}$ distribution for 50 MFIM capacitors with $T_{FE}$=10nm at different $V_{SET}$ and (b) standard deviation (σ) and mean (μ) of $P_{R+}$ (σ($P_{R+}$) and μ($P_{R+}$) respectively) versus $V_{SET}$ for different $T_{FE}$ (=5, 7, and 10nm).

its low gradient energy [33], is another crucial factor dictating variations due to its random nature (especially domain nucleation) [35]. Based on the phase-field simulations, the works in [34], [35], [41] have shown that *P* switching mechanisms (domain nucleation and domain growth) and domain characteristics are $V_{SET}$ and $T_{FE}$-dependent, which causes $P_R$ variations to depend on $V_{SET}$ and $T_{FE}$ strongly. Despite these insights from simulations, experimental validation of these trends, especially with regard to $T_{FE}$, is limited. Moreover, exploring the correlation of variations with factors (like domain nucleation, polycrystallinity, and traps) is crucial for advancing the understanding of device-to-device variations in HZO devices.

This work experimentally characterizes the variations in $P_R$ as a function of $V_{SET}$ and $T_{FE}$ and provides evidence that aligns with previous simulation-based analyses and insights. By measuring $P_R$ from polarization-voltage (*P-V*) characteristics across 50 MFIM capacitors, each for three different $T_{FE}$, we capture the relation between $P_R$, $V_{SET}$, and $T_{FE}$. The key contributions of this work are summarized below.

- We experimentally establish that the standard deviation of $P_R$ (σ($P_R$)) exhibits a non-monotonic trend with respect to $V_{SET}$, peaking at $V_{SET}$ near $V_C$.

- Our measurement results reveal that the σ($P_R$) depends on $T_{FE}$, exhibiting distinct trends across low-, mid-, and high-$V_{SET}$ ranges.

- By analyzing the correlation between $P_R$, $V_C$, and saturation polarization ($P_S$) measured across different $V_{SET}$ and $T_{FE}$, we discuss the role of polycrystallinity, charge trapping effect, and domain nucleation in dictating the $P_R$ variations.

- We support our experimental findings with the interpretations from our in-house 3D phase-field model.

## II. MFIM Capacitor Fabrication and Measurements

We fabricate HZO-based MFIM capacitors comprising 70nm tungsten (W) for the top and bottom metal electrodes, a ferroelectric HZO layer with various thicknesses ($T_{FE}$=5, 7, 10nm), and a 3nm $Al_2O_3$ insulator or dielectric (DE) layer, as shown in Fig. 1(a). First, the W bottom electrode is deposited on a bare Si wafer using sputtering. Atomic layer deposition (ALD) is then utilized to deposit HZO and $Al_2O_3$ layers at 200ºC. Following this, W is sputtered to form the top electrode. This is followed by 30s of rapid thermal annealing (RTA) at 300ºC in the nitrogen ($N_2$) atmosphere. The capacitor area is defined by the top electrode, patterned as a circle with a 25μm radius through the lift-off process using photolithography.

The *P-V* measurements of 50 fabricated MFIM capacitors at each $T_{FE}$ are performed using a Radiant Premier II Ferroelectric Tester through a probe station. The applied voltage ($V_{app}$) is swept at 10kHz frequency from a fixed $V_{RESET}$ to a different $V_{SET}$ (Fig. 1(b)). Note that this biasing scheme is consistent with the programming methodology used in previous experimental works such as [27]. For consistency, $V_{RESET}$ for each $T_{FE}$ is selected to obtain a similar magnitude of average electric field (*E*-field) across the FE layer at $V_{RESET}$ (=-10, -8.3, and -7.1V for $T_{FE}$=10, 7, and 5nm). (Note: The *E*-field is estimated by using phase-field simulations). To minimize the effect of cycle-to-cycle variations [34], the *P-V* curves used in the analysis are averaged over five measurement cycles. Fig. 1(c) illustrates the *P-V* plots averaged for the 50 MFIM capacitors with $T_{FE}$=10nm for various $V_{SET}$. All these curves start from negative $P_S$ ($P_{S-}$) at the fixed $V_{RESET}$, pass negative $P_R$ ($P_{R-}$) at $V_{app}$=0V, and reach varying *P* values depending on $V_{SET}$. After that, *P* returns to $P_{S-}$ at $V_{RESET}$ as $V_{app}$ is swept backward from $V_{SET}$. The *P-V* curves reach positive $P_S$ ($P_{S+}$) only for the major loop, where |$V_{SET}$|=|$V_{RESET}$|. The positive $P_R$ ($P_{R+}$) is measured at $V_{app}$=0V on the return path from $V_{SET}$ to $V_{RESET}$ for each cycle. $V_C$ is extracted as the voltage where *P*=0 on the path from $V_{RESET}$ to $V_{SET}$ in the major loop. Note that we consider the scenario where different $V_{SET}$ values are applied to achieve multiple states of the MFIM from an initial fixed reset state (spontaneous state after the application of fixed $V_{RESET}$). Therefore, we characterize the standard deviation of $P_{R+}$ (σ($P_{R+}$)) to analyze the device-to-device variations in $P_R$. Due to the symmetry of *P-V* characteristics, analyzing $P_{R-}$ for different $V_{RESET}$ (after the application of fixed $V_{SET}$ application) is equivalent to our analysis. Here, we focus on the trends in $P_{R+}$ variations.

## III. Experimental Results

### A. The Non-monotonic Trend of $P_{R+}$ variations with $V_{SET}$

Fig. 2(a) shows the measured $P_{R+}$ distributions for 50 MFIM capacitors with $T_{FE}$=10nm at various $V_{SET}$ values revealing a significant dependence of $P_{R+}$ variations on $V_{SET}$ (Fig. 2(b)). $\sigma(P_{R+})$ shows a non-monotonic behavior with respect to $V_{SET}$ (as also predicted via simulations in [35]). Initially, the distribution shows a mild reduction in $\sigma(P_{R+})$ as $V_{SET}$ increases from 4.0V to 5.3V (region-1 or low-$V_{SET}$ range). The distribution substantially increases its spread as $V_{SET}$ increases from 5.3V to 6.2V (region-2 or mid-$V_{SET}$ range). As $V_{SET}$ increases further, $\sigma(P_{R+})$ reduces (region-3 or high-$V_{SET}$ range). It can be observed in Fig. 2(b) that $\sigma(P_{R+})$ exhibits a peak at a particular $V_{SET}$, which is annotated as $V_{peak}$. Given that the mean $P_{R+}$ ($\mu(P_{R+})$) at $V_{peak}$ is close to 0 (Fig. 2(c)), it follows that $\sigma(P_{R+})$ reaches a peak point around $V_{SET}{\sim}V_C$.

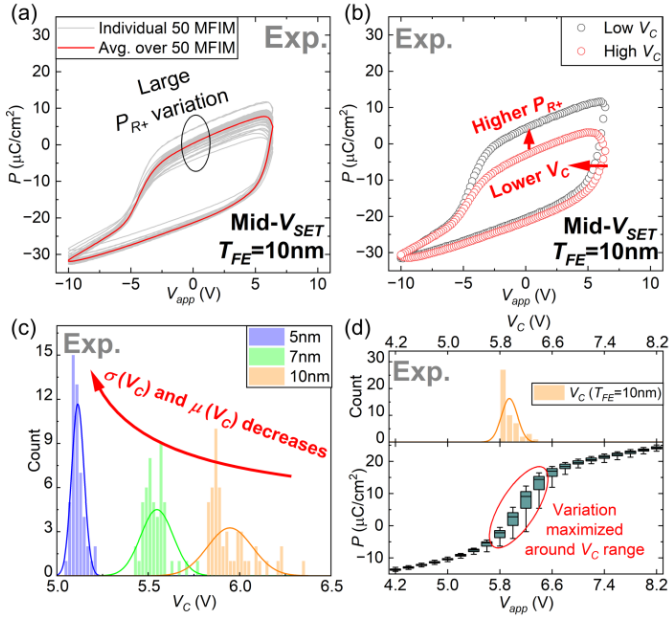

Fig. 3. (a) *P-V* curves of individual and averaged 50 MFIM capacitors with $T_{FE}$=10nm at $V_{SET}$=6.4V (b) *P-V* minor loops ($V_{SET}$=6.4V) with different $V_C$. At $V_{SET}$ range around $V_C$, $P_{R+}$ is largely dependent on $V_C$. (c) Measured $V_C$ distribution from 50 MFIM capacitors of different $T_{FE}$. (d) Measured $V_C$ distribution (top) and box plot depicting the $P$ variations around $V_C$ (bottom) of 50 MFIM with $T_{FE}$=10nm.

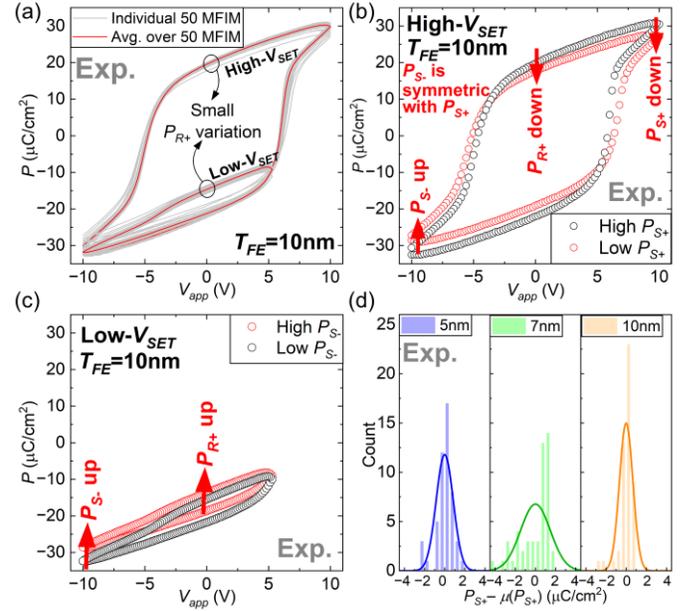

Fig. 4. (a) *P-V* curves of individual and averaged 50 MFIM capacitors at $V_{SET}$=10.0V and $V_{SET}$=5.9V (b) *P-V* major loops ($V_{SET}$=10.0V) with different $P_{S+}$. (c) *P-V* minor loops ($V_{SET}$=5.9V) with different $P_{S-}$. In the high-$V_{SET}$ range, $P_{R+}$ is largely dependent on $P_{S+}$, and $P_{S-}$ is symmetric with $P_{S+}$. (d) The distribution of $P_{S+} - \mu(P_{S+})$ from 50 MFIM capacitors of different $T_{FE}$. The $\sigma(P_{R+})$ trends in Fig. 2(b) at the low and high-$V_{SET}$ range follow that of the $P_{S+}$ distribution.

With decreasing $T_{FE}$, the non-monotonicity of $\sigma(P_{R+})$ becomes less prominent. The peak value of $\sigma(P_{R+})$ decreases as $T_{FE}$ scales down, reducing from 1.73μm/cm² at $T_{FE}$=10nm to 1.11μm/cm² (0.73×) and 0.71μm/cm² (0.41×) at $T_{FE}$=7 and 5nm. To understand this reduction, we plot the measured $V_C$ distributions in Fig. 3(c). We observe a decrease in $V_C$ variations as $T_{FE}$ decreases. Since $P_{R+}$ variations are largely influenced by $V_C$ in the mid-$V_{SET}$ range (Fig. 3(b)), $P_R$ variations also decrease with $T_{FE}$ scaling. Compared to $T_{FE}$=10nm, $\sigma(V_C)$ reduces to 0.72× at $T_{FE}$=7nm and 0.28× at $T_{FE}$=5nm. Therefore, we can correlate a decrease in the peak $\sigma(P_{R+})$ primarily with lower $V_C$ variations at a scaled $T_{FE}$. Also, the $V_C$ distributions explain the $V_{peak}$ shift towards lower $V_{SET}$. The mean value of $V_C$ ($\mu(V_C)$) shows a shift to a lower value as $T_{FE}$ is scaled (6.0V→5.5V→5.1V for $T_{FE}$=10, 7, and 5nm). This is primarily due to an increase in E-field for lower $T_{FE}$ at a given $V_{app}$. This aligns with $V_{peak}$ shifts shown in Fig. 2(b) (6.4V→5.7V→5.0V for $T_{FE}$=10, 7, and 5nm). These results illustrate that $V_{peak}$ is linked to $V_C$ and decreases as $T_{FE}$ is scaled.

However, $\sigma(P_{R+})$ in low- and high-$V_{SET}$ regions does not display any trend depending on $T_{FE}$ (unlike in mid-$V_{SET}$). In these regions, $\sigma(P_{R+})$ is the highest for $T_{FE}$=7nm, followed by 5nm and 10nm for both low- and high-$V_{SET}$ regions (Fig. 2(b)).

To understand the trend in the high-$V_{SET}$ regions, let us look at Fig. 4(b), which points to a relationship between $P_{S+}$ variations and $P_{R+}$ variations. Specifically, MFIM samples with lower $P_{S+}$ are likely to exhibit lower $P_{R+}$. Measured $P_{S+}$ distributions for each $T_{FE}$ in Fig. 4(d) show that the variation is the largest for $T_{FE}$=7nm, followed by 5nm and 10nm, reflecting the trend observed in $P_{R+}$ variations in the high-$V_{SET}$ region.

In the low-$V_{SET}$ region, two key factors dictate $P_{R+}$ variations. First, lower (more negative) $P_{S-}$ results in a lower $P_{R+}$, as shown in Fig. 4(c), indicating that the variations in $P_{S-}$ directly drive $P_{R+}$ variations. Second, because $P_{S-}$(<0) and $P_{S+}$(>0) exhibit symmetry–i.e., more negative $P_{S-}$ implies higher $P_{S+}$ (Fig. 4(b)) –a negative correlation exists between $P_{S-}$ and $P_{S+}$. As a result, a negative correlation is established between $P_{S+}$ and $P_{R+}$, where a larger value of $P_{S+}$ is associated with a smaller value of $P_{R+}$ (due to a large negative value of $P_{S-}$). Furthermore, the consistency in the trend of $\sigma(P_{R+})$ and $\sigma(P_{S+})$ with respect to $T_{FE}$ (Fig. 4(d)) substantiates the influence of $P_{S+}$ variations on $P_{R+}$ variations in the low-$V_{SET}$ region.

### B. Correlation Analysis for $P_{R+}$ with $V_C$ and $P_{S+}$

To further understand the relationships between $P_{R+}$, $V_C$, and $P_{S+}$, we present scatter plots in Fig. 5(a) showing the correlation of measured values of $P_{R+}$ with $P_{S+}$ and $V_C$ across different $V_{SET}$ values for 50 MFIM capacitors at $T_{FE}$=10nm. At low $V_{SET}$ (=4.0V), $P_{R+}$ shows a strong correlation with $P_{S+}$, suggesting that $P_{R+}$ variations at the low-$V_{SET}$ range are mainly driven by factors influencing $P_{S+}$. As $V_{SET}$ approaches the mid-$V_{SET}$ range (close to $V_C$), the correlation between $P_{R+}$ and $P_{S+}$ becomes weaker, showing a minimal correlation at $V_{SET}$=6.2V. With a further increase in $V_{SET}$ (high-$V_{SET}$ range), the correlation between $P_{R+}$ and $P_{S+}$ becomes strong again. Notably, $P_{R+}$ negatively correlates with $P_{S+}$ at the low $V_{SET}$ range (4.0V) while positively at the high $V_{SET}$ range (10.0V). At high-$V_{SET}$,

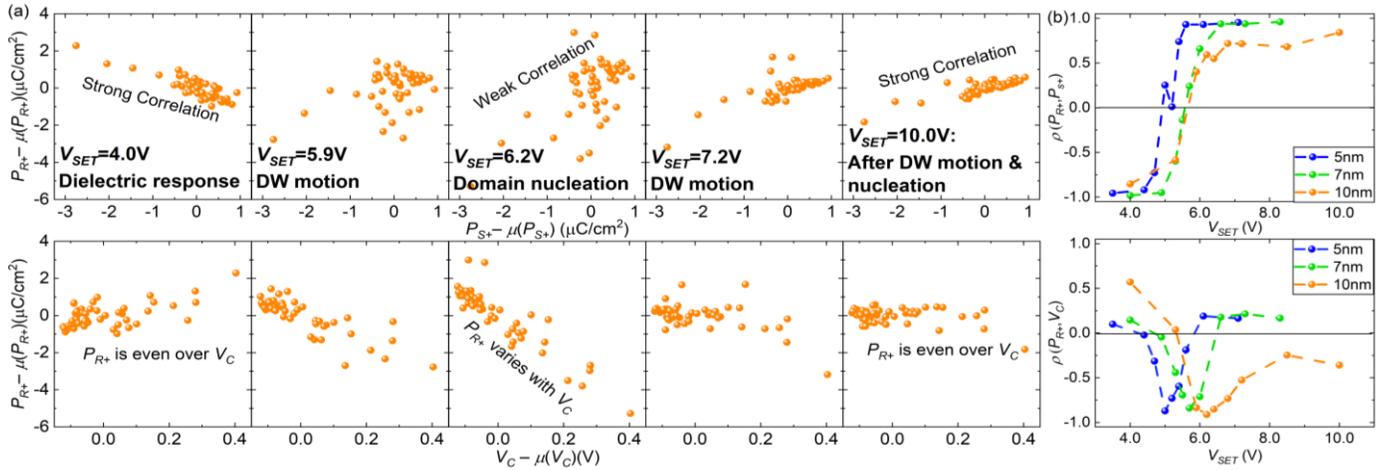

Fig. 5. (a) Measured correlation plots of $P_{R+} - \mu(P_{R+})$ with $P_{S+} - \mu(P_{S+})$ (up) and $V_C - \mu(V_C)$ (down) of the 50 MFIM capacitors with $T_{FE}$=10nm and (b) the correlation coefficient of $P_{R+}$ with $P_{S+}$ ($\rho(P_{S+}, P_{R+})$) (up) and with $V_C$ ($\rho(V_C, P_{R+})$) (down) as a function of $V_{SET}$ for different $T_{FE}$. At the low and high $V_{SET}$ (4.0 and 10.0V), the effect of polycrystallinity is stronger, while domain nucleation has a huge effect at the middle $V_{SET}$ (6.2V).

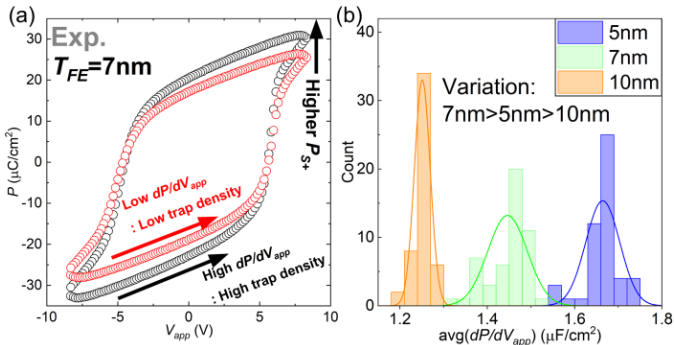

Fig. 6. (a) P-V loops with different trap densities. The black P-V loop features higher $dP/dV_{app}$ and $P_{S+}$ than the red one representing higher trap density. (b) the distribution of $dP/dV_{app}$ for different $T_{FE}$. The order of $dP/dV_{app}$ variations for different $T_{FE}$ is consistent with that of $P_{S+}$ variations shown in Fig. 4(d)

MFIM with higher $P_{S+}$ yields higher $P_{R+}$, as discussed in Section III. A, leading to a positive correlation. On the other hand, in the low-$V_{SET}$ range, $P_{R+}$ is mainly determined by $P_{S-}$–i.e., higher $P_{S-}$ (less negative $P_{S-}$) yields higher $P_{R+}$. Moreover, recalling the symmetry between $P_{S-}$ and $P_{S+}$, a higher (less negative) $P_{S-}$ implies a lower $P_{S+}$. As discussed before, by combining these two correlations, we can deduce that lower $P_{S+}$ results in higher $P_{R+}$ in the low-$V_{SET}$ range. This results in a negative correlation at $V_{SET}$=4.0V.

Now, let us discuss the correlation of $P_{R+}$ with $V_C$. In the low-$V_{SET}$ range ($V_{SET}$=4.0V), $P_{R+}$ shows minimal correlation with $V_C$. As $V_{SET}$ increases, approaching 6.2V (mid-$V_{SET}$), a strong negative correlation emerges showing that capacitors with low $V_C$ undergo higher $P$ switching at a given $V_{app}$. In the high-$V_{SET}$ range, the correlation becomes weak again, indicating a reduced influence of $V_C$ on $P_{R+}$.

The correlation coefficient ($\rho$) of $P_{R+}$ with $P_{S+}$ ($\rho(P_{R+},P_{S+})$) and $V_C$ ($\rho(P_{R+},V_C)$) across different $V_{SET}$ values (shown in Fig. 5(b)) summarize the discussion above. Initially, $\rho(P_{R+},P_{S+})$ is around -1, indicating a strong negative correlation. It approaches 0 at $V_{SET}$~$V_C$ and reaches +1 at the high-$V_{SET}$ values. For $\rho(P_{R+}, V_C)$, the coefficient dips to almost -1 around $V_C$, representing a substantial negative correlation, and remains closer to 0 at other $V_{SET}$ values. We also observe that these trends hold for all $T_{FE}$ (with some differences that will be discussed subsequently).

## IV. ANALYSIS OF $P_{R+}$ VARIATIONS: MECHANISMS FOR DIFFERENT $V_{SET}$ RANGE

### A. $P_{R+}$ Variation Mechanism in the Low- and High-$V_{SET}$ Regions

Here, we focus on three main mechanisms for $P_{R+}$ variations: MD $P$ switching, polycrystallinity, and the charge trap effects at the FE-DE interface. As we will discuss, MD $P$ switching is not a major contributor to the variation in the low- and high-$V_{SET}$ range due to small $P$ switching in the former and nearly complete $P$ switching in the latter region. Therefore, the primary influence on the variations comes from charge trapping [36], [37], [38] and polycrystallinity [34]. The strong correlation between $P_{S+}$ and $P_{R+}$ observed experimentally confirms that the $P_{R+}$ variations are mainly driven by $P_{S+}$ variations (Fig. 4(b) and (c)) originating from charge trap effects and polycrystallinity. Charge trapping affects $P_{S+}$ variations through two mechanisms. First, trapped charges directly influence $P$ within the FE layer [36], [37]. Second, trapped charges modify the depolarization field across the FE layer [42], resulting in a change in the DW density [41].

Apart from charge trapping, polycrystallinity is another key contributor to the $P_{S+}$ variations. In polycrystalline HZO, the grain orientation angle ($\theta$)–the angle between the $P$ direction (c-axis of the orthorhombic unit cell) and the physical out-of-plane axis–governs $P$ switching. Since the component of applied out-of-plane $E$-field that is parallel to $P$ contributes to switching [34], variations in $\theta$ among grains lead to a non-uniform $P$ switching. Specifically, grains with larger $\theta$ are stimulated with a weaker effective $E$-field, which leads to higher $V_C$. Due to the resultant suppressed $P$ switching, a fraction of the original $P$ domains remains even after the $P$ switching in the major loop, leading to a reduction in $P_{S+}$ for higher $\theta$ [34].

To investigate the impact of charge trapping, we analyze $dP/dV_{app}$ in voltage regions with minimal $P$ switching and dominant dielectric response. Previous studies have shown that trapped charges at the FE-DE interface directly affect $P$ [36], resulting in different $dP/dV_{app}$ in the non-switching region [37]. Meanwhile, polycrystallinity has little impact on $dP/dV_{app}$ since the permittivity of FE, which determines $dP/dV_{app}$, remains mostly invariant irrespective of grain orientation [34]. The measured variations in $dP/dV_{app}$ ($\sigma(dP/dV_{app})$) across 50 MFIM capacitors follow the order $T_{FE}$=7nm>5nm>10nm (Fig. 6(b)), aligning with the trends observed for $\sigma(P_{S+})$ in Fig. 4(d). This correlation between $\sigma(dP/dV_{app})$ and $\sigma(P_{S+})$ highlights a significant role of charge trapping in governing $P_{S+}$ variations, which in turn contribute to $P_{R+}$ variations in the low- and high-$V_{SET}$ ranges.

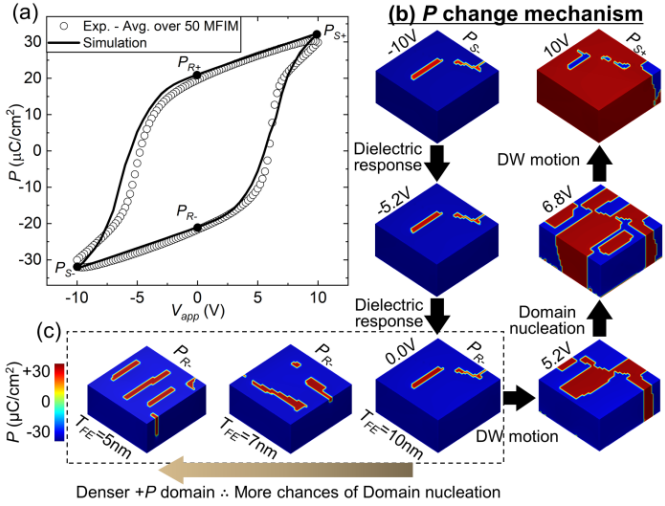

Fig. 7. (a) Simulated (line) and experimental (symbol) *P-V* with $T_{FE}$=10nm. The experimental *P-V* is averaged over 50 MFIM capacitors (b) *P* maps from the phase-field simulation illustrating the *P* changing mechanism as $V_{app}$ increases from -10V ($V_{RESET}$) to 10V ($V_{SET}$) and (c) simulated *P* maps of $P_{R-}$ for different $T_{FE}$ (=5, 7, and 10nm)

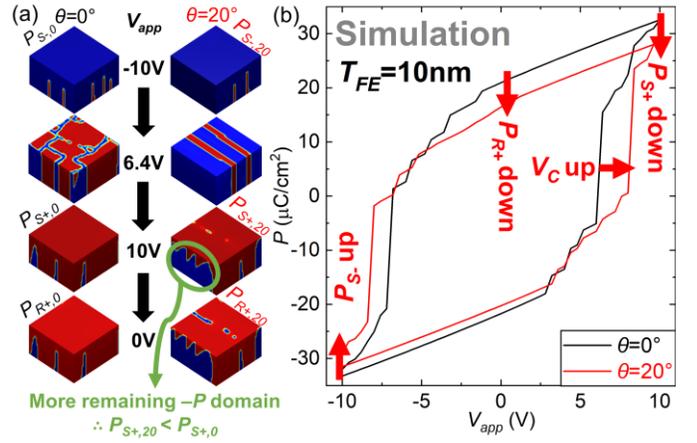

Fig. 8. (a) *P* maps during the $V_{app}$ sweep (-10V($V_{RESET}$)→10V($V_{SET}$)→0V) and (b) *P-V* loops of HZO samples with $\theta$=0° and 20° from the phase-field simulations. $P_{S-,\theta}$, $P_{S+,\theta}$, and $P_{R+,\theta}$ represent $P_{S-}$, $P_{S+}$, and $P_{R+}$ of the grain with grain orientation angle $\theta$.

It may be noted that the impact of polycrystallinity becomes more pronounced as $T_{FE}$ increases. It is well known that a larger $T_{FE}$ exhibits a larger average grain size [30], reducing the number of grains ($N_{grain}$) within a fixed cross-sectional area. As $N_{grain}$ reduces, statistical averaging becomes less effective, resulting in larger device-to-device variations in $\theta$. Thus, polycrystalline-induced $P_{S+}$ variations increase, which indicates the stronger influence of polycrystallinity for MFIM with larger $T_{FE}$.

In our experiments, the effect of traps on $P_S$ and $P_R$ variations is the largest for $T_{FE}$=7nm, as discussed before, while polycrystallinity-induced variations are expected to be the largest for $T_{FE}$=10nm. Based on the trends of $\sigma(P_{S+})$ and $\sigma(P_{R+})$ in the low- and high-$V_{SET}$ regions, it appears that the former effect is more dominant in our experiments. This can be attributed to the large cross-sectional area of our experimental MFIM samples ($\approx1.96\times10^3\mu m^2$), which makes the impact of $N_{grain}$ on variations relatively less significant. However, for scaled devices, polycrystallinity can be a significant contributor, as has been shown in previous simulation studies [34]. Despite the large cross-sectional area, polycrystallinity does show some effects on the trends of the correlations between $P_{R+}$ and $V_C$. Specifically, the $\rho(P_{R+},V_C)$ trend differs significantly for $T_{FE}$=10nm, where the effect of polycrystallinity on variations is the strongest. As increased $\theta$ leads to lower $P_{S+}$ and higher $V_C$, these two parameters exhibit a non-zero correlation. Consequently, at high-$V_{SET}$, $T_{FE}$=10nm displays negative $\rho(P_{R+},V_C)$ since $\rho(P_{R+},P_{S+})$ = +1 and $\rho(P_{S+},V_C)$ is negative. This is unlike 7nm and 5nm, where $\rho(P_{R+},V_C)\approx0$, which indicates a lower effect of polycrystallinity (and $\theta$). At low $V_{SET}$, $T_{FE}$=10nm shows a positive $\rho(P_{R+},V_C)$ (0.569), significantly higher than that of $T_{FE}$=7nm (0.145) and 5nm (0.098). Again, this is due to the fact that at $T_{FE}$=10nm, $\rho(P_{S+},V_C)$ is negative (due to polycrystallinity) and $\rho(P_{R+},P_{S+})$ is negative (as discussed above), which yields a significantly positive $\rho(P_{R+},V_C)$. Fig. 5(b) illustrates this trend, suggesting an increased impact of polycrystallinity in $P_{R+}$ variations for larger $T_{FE}$ (=10nm).

### B. $P_{R+}$ Variation Mechanisms in the Mid-$V_{SET}$ Region

In the mid-$V_{SET}$ range, $P_{R+}$ variations are mainly dictated by $V_C$ variations as shown in the correlation plots (Fig. 5(a)). The sharp *P* switching in the mid-$V_{SET}$ amplifies $V_C$-induced $P_{R+}$ variations (Fig. 4(a) and (d)), which leads to the $\sigma(P_{R+})$ peaks as shown in Fig. 2(b). According to the previous simulation-based studies [34], [35], the strong influence of domain nucleation around the mid-$V_{SET}$ range leads to large variations in $P_{R+}$. Domain nucleation features inherent randomness and sharp *P* switching (a large amount of *P* switched in a small voltage range). The randomness of domain nucleation arises from the polycrystallinity of HZO and random MD *P* configurations [34], [35]. This leads to different samples nucleating at different $V_{app}$, represented by the $V_C$ variations (Fig. 3(c)). Due to the sharp *P* switching associated with domain nucleation, a subset of samples with lower $V_C$ (early nucleating samples) reaches a higher *P* at a given $V_{app}$ [35]. As shown in Fig. 3(d), this effect broadens the *P* distribution around $V_C$ due to the sharp *P* switching, leading to large $P_{R+}$ variations (Fig. 3(a)) as $V_{app}$ returns from $V_{SET}\sim V_C$ to 0V.

The impact of $T_{FE}$ on $\sigma(P_{R+})$ peaks is consistent with the attribution of polycrystallinity and MD effect on $P_{R+}$ variations. In a thinner FE layer, the -*P* domain patterns become denser to compensate for increased depolarization energy [41]. This increased domain density limits the room for domain nucleation and thus suppresses it. Therefore, DW motion becomes a prevalent *P* switching mechanism in a thinner FE layer, leading to reduced randomness and, consequently, lower $V_C$ (Fig. 3(c)) and $P_{R+}$ variations (Fig. 2(b)). Moreover, reduced statistical averaging in $\theta$ due to small $N_{grain}$ at large $T_{FE}$ also contributes to larger variations in MFIM capacitors with a thicker FE layer (especially in scaled devices).

### C. Analysis of Factors Affecting $P_{R+}$ Variations Using 3D Phase-field Model

To gain further insights into the mechanisms governing the $P_{R+}$ variations, we utilize a calibrated 3D phase-field model [34], [35] to simulate the MFIM structure. Our 3D multi-grain phase-field framework calculates the $P$ profile in an MFIM stack by solving time-dependent Ginzburg Landau (TDGL) and Poisson's equations. The model includes the polycrystalline nature of HZO by associating the 3D grain-growth model with $\theta$ (details in [34], [35]). It should be noted that the trap effect is not considered explicitly; instead, it is abstracted within calibrated parameters in the model. Thus, this model-based analysis shed light only on the MD and polycrystallinity effects on the $P_{R+}$ variations. The black solid line in Fig. 7(a) represents the calibrated $P$-$V$ curve from the model for $T_{FE}$=10nm, which closely matches the average of measured $P$-$V$ curves over 50 MFIM capacitors (black symbols). The $P$ profiles from the phase-field simulation (Fig. 7(b)) illustrate the voltage dependence of $P$ switching mechanisms in the FE layer.

As expected, domain nucleation is observed at the $V_{app}$ around the mid-$V_{SET}$ range (5.2-6.8V), which explains the significant variations observed in this region. The model also demonstrates the reduced chance of domain nucleation at smaller $T_{FE}$ due to denser +$P$ domains at $P_{R-}$ (Fig. 7(c)) [41]. These results are consistent with experimental data indicating the significant impact of domain nucleation on $P_{R+}$ variations in the mid-$V_{SET}$ range and its dependence on $T_{FE}$. In low- and high-$V_{SET}$ ranges, dielectric response and domain wall (DW) motion are prevalent mechanisms. As the randomness and magnitude in $P$ switching reduces, the role of the MD effect in dictating $P_{R+}$ variations becomes marginal.

Our model suggests polycrystallinity plays a key role in $P_{R+}$ variations in the low and high-$V_{SET}$ regions (Recall, traps are not included in our model). At $V_{app}$=10V, more -$P$ domains are observed in the grain with $\theta$=20° (compared to $\theta$=0°), leading to lower $P_{S+}$ (Fig. 8(b)). Since the dielectric response is prevalent during the backward $V_{app}$ sweep from 10V to 0V (Fig. 8(a)), lower $P_{S+}$ directly leads to lower $P_{R+}$. Likewise, in the low-$V_{SET}$ range, polycrystallinity-induced $P_{S-}$ variations lead to $P_{R+}$ variations. Furthermore, our model exhibits increased $V_C$ in the grain with $\theta$=20° as discussed before. Therefore, our model shows results consistent with the discussions above confirming that grains with larger $\theta$ exhibit smaller $P_{S+}$ and increased $V_C$.

## V. CONCLUSION

We experimentally investigated the mechanisms of the $P_R$ variations in different $V_{SET}$ regions and $T_{FE}$ for HZO-based MFIM capacitors, confirming the previously predicted non-monotonic behavior of $\sigma(P_{R+})$ with $V_{SET}$. By analyzing the correlation between measured $P_R$, $V_C$, and $P_S$ across different $V_{SET}$, we highlight which factor has more effect on $P_R$ variations. $P_{S+}$ variation mainly drives $P_{R+}$ variation in the low- and high-$V_{SET}$ regions, originating from charge trap at the FE-DE interface and polycrystallinity effects. Specifically, the alignment between the trends of $dP/dV_{app}$ and $P_{S+}$ as a function of $T_{FE}$ emphasizes the dominant impact of charge trapping on $P_{R+}$ variation. Additionally, the polycrystallinity effect impacts the $P_{R+}$ variation at larger $T_{FE}$ (=10nm) due to reduced $N_{grain}$, which reduces the statistical averaging effect. Conversely, near the mid-$V_{SET}$ region, $P_{R+}$ variation is primarily dictated by $V_C$ variation resulting from MD $P$ switching combined with polycrystallinity. The random nature and sharp $P$ switching amplify the $P_{R+}$ variations, which leads to peaks in $\sigma(P_{R+})$. This effect diminishes as $T_{FE}$ decreases due to suppressed domain nucleation as well as reduced polycrystallinity effect. Our 3D phase-field model substantiates the analysis from experimental data. The model offers insights into the $P$ change mechanism for different voltage and domain density trends with respect to $T_{FE}$ and the role of polycrystallinity. Through our analysis based on experiments and simulations, we advance the understanding of $P_R$ variations and the underlying mechanisms.